\documentclass[prd,aps,twocolumn,showpacs,preprintnumbers,amsmath,amssymb,nofootinbib]{revtex4}
\usepackage[dvips]{graphicx}
\usepackage{epsf}
\usepackage{amsmath}
\usepackage{amssymb}

\usepackage{graphicx}
\usepackage{dcolumn}
\usepackage{bm}
\begin{document}

\title{Deep connection between $f(R)$ gravity and the interacting dark sector model }

\author{Jian-Hua He$^{1}$, Bin Wang$^{1}$, Elcio Abdalla$^{2}$}
\affiliation{
$^{1}$ INPAC and Department of Physics, Shanghai Jiao Tong University,
Shanghai 200240, China\\
$^{2}$ Instituto de Fisica, Universidade de Sao Paulo, CP 66318, 05315-970, Sao Paulo, Brazil}

\begin{abstract}
We examine the conformal equivalence between the $f(R)$ gravity and the interacting dark sector model. We  review the well-known result that the conformal transformation physically corresponds to the mass dilation which marks the strength of interaction between dark sectors. Instead of modeling f(R) gravity in the Jordan frame, we construct the $f(R)$ gravity in terms of mass dilation function in the Einstein frame. We find that the condition to keep  $f(R)$ gravity consistent with CMB observations ensures the energy flow from dark energy to dark matter in the corresponding interacting model, which meets the requirement to alleviate the coincidence problem in the Einstein framework.

\end{abstract}
\pacs{98.80.Cq}

\maketitle

\section{Introduction}
There are concordance pictures indicating that our universe is experiencing an accelerated expansion.
This acceleration is believed to be driven by a yet unknown dark energy (DE) in the framework of Einstein gravity.  The leading interpretation of such a DE is the cosmological constant. However  the cosmological constant falls far below the value predicted by any sensible quantum field theory and it unavoidably leads to the coincidence problem, namely, "why are the vacuum and matter energy densities of precisely the same order today?".

Considering that DE and dark matter (DM) contribute significant fractions of the contents of the universe, it is natural, in the framework of field theory, to consider the interaction between them.  The possibility that DE and DM can interact has been studied extensively recently \cite{1}-\cite{28}. It has been shown that the coupling between DE and DM can provide a mechanism to alleviate the coincidence problem \cite{2}-\cite{6}\cite{19}\cite{199}.  Complementary Observational signatures of the interaction between DE and DM have been obtained from the cosmic expansion history by using the WMAP, SNIa, BAO and SDSS data etc \cite{16}-\cite{20} as well as the growth of cosmic structure \cite{22}-\cite{28}.

Another possible way to explain the acceleration of the universe is to modify Einstein gravity. One of the attempts is called the $f(R)$ gravity, in which the Lagrange density $f$ is an arbitrary function of $R$ \cite{30}-\cite{33}. $f(R)$ gravity is considered as the simplest modification of Einstein's general relativity. However, it is quite non-trivial to construct a viable $f(R)$ model satisfying both cosmological and local gravity constraints \cite{34}-\cite{AmendolaPRL}. It is possible to transform the action of $f(R)$ gravity from the original Jordan frame to the Einstein frame by using conformal transformations \cite{tsujikawa}. The $f(R)$ gravity turns out to be conformally equivalent to an interacting model of DE and DM. In the Einstein frame, the model does not possess a standard matter-dominated epoch as in the Jordan frame, but contains the coupling between the canonical scalar field to the non-relativistic matter.

In this work we will further illustrate the well-known result of conformal equivalence between the $f(R)$ gravity and the model of interacting dark sectors which is addressed in many papers ~\cite{magnano,tsujikawa}.  We will construct the $f(R)$ models in terms of the mass dilation rate which describes the strength of the interaction between dark sectors.  With the coupling strength, it is easy to construct the viable $f(R)$ model realizing a reasonable cosmic expansion history. Furthermore, we will show that the condition that $f(R)$ gravity avoids the short-timescale instability and maintains the agreement with CMB is exactly equivalent to the requirement of an energy flow from DE to DM in the interaction model to ensure the minimization of the coincidence problem in the Einstein frame \cite{19}\cite{199}.

In the following section we will first present the general formalism of $f(R)$ gravity and its conformal description in the Einstein frame. We will relate the conformal transformation to the concept of mass dilation function $\Gamma$. In section~\ref{II}, we apply the conformal discussion to cosmology. We show again the equivalence between the $f(R)$ gravity and the conformal viable cosmological model of interaction between dark sectors and  we also go back to the Jordan frame to check the consistency. In the last section we present our summary and discussion.

\section{f(R) gravity and conformal transformation\label{I} }
We start with the 4-dimensional action in $f(R)$ gravity in the Jordan frame
\begin{equation}
S=\frac{1}{2\kappa^2}\int d^4x\sqrt{-g}f(R)+\int d^4x\mathcal{L}^{(m)}\quad,\label{action}
\end{equation}
where $R$ is the Ricci scalar, $\kappa^2=8\pi G$, and $\mathcal{L}^{(m)}$ is the matter Lagrangian. Variation with respect to the metric $g_{\mu\nu}$ yields the field equation
\begin{equation}
FR_{\mu\nu}-\frac{1}{2}fg_{\mu\nu}-\nabla_{\mu}\nabla_{\nu}F+g_{\mu\nu}\Box F=\kappa^2T_{\mu\nu}^{(m)}\quad ,\label{Einstein}
\end{equation}
where $F=\frac{\partial f}{\partial R}$. When $f(R)=R$, the above equation is just the Einstein equation. The field equation can be rewritten in the form
\begin{equation}
G_{\mu\nu}=\frac{1}{2F}(f-RF)g_{\mu\nu}+\frac{1}{F}(\nabla_{\mu}\nabla_{\nu}-g_{\mu\nu}\Box)F+\kappa^2\frac{T_{\mu\nu}^{(m)}}{F}\quad ,
\end{equation}
where $G_{\mu\nu}=R_{\mu\nu}-\frac{1}{2}Rg_{\mu\nu}$. Defining the effective energy-momentum tensor $T_{\mu\nu}^{(e)}$ as
\begin{equation}
T_{\mu\nu}^{(e)}=\frac{1}{2\kappa^2}(f-RF)g_{\mu\nu}+\frac{1}{\kappa^2}(\nabla_{\mu}\nabla_{\nu}-g_{\mu\nu}\Box)F\quad,
\end{equation}
we can recast the field equation into ~\cite{tsujikawa}
\begin{equation}
G_{\mu\nu}=\frac{\kappa^2}{F}(T_{\mu\nu}^{(e)}+T_{\mu\nu}^{(m)})\quad\label{p} .
\end{equation}

It is possible to discuss the $f(R)$ gravity in the Einstein frame under the conformal transformation \cite{dicke}
\begin{equation}
\tilde{ds}^2=\Omega^2ds^2,\quad,\tilde{g}_{ab}=\Omega^2g_{ab},\quad \tilde{g}^{ab}=g^{ab}/\Omega^2
\end{equation}
where $\Omega^2$ is a positive defined conformal factor and a tilde represents quantities in the Einstein frame.

As explained in the appendix, the Einstein tensor $G_{\mu\nu}$ transforms into
\begin{eqnarray}
G_{\mu\nu}&=&\tilde{G}_{\mu\nu}+2\tilde{\nabla}_{\mu}\omega\tilde{\nabla}_{\nu}\omega+2\tilde{\nabla}_{\mu}\tilde{\nabla}_{\nu}\omega\nonumber\\
&-&2\tilde{g}_{\mu\nu}\tilde{\Box}\omega+\tilde{g}_{\mu\nu}\tilde{g}^{\tau\sigma}\tilde{\nabla}_{\tau}\omega\tilde{\nabla}_{\sigma}\omega\quad,\label{G}
\end{eqnarray}
where $\omega=\ln \Omega$.

The effective energy-momentum
tensor $T_{\mu\nu}^{(e)}$ transforms as
\begin{eqnarray}
T_{\mu\nu}^{(e)}&=&\frac{f-RF}{2\kappa^2\Omega^2}\tilde{g}_{\mu\nu}\nonumber\\
&+&\frac{1}{\kappa^2}\left(\tilde{\nabla}_{\mu}\tilde{\nabla}_{\nu}F-\tilde{g}_{\mu\nu}\tilde{\Box}F\right)\nonumber \\
&+&\frac{1}{\kappa^2}\left(2\tilde{\nabla}_{(\mu}F\tilde{\nabla}_{\nu)}\omega+\tilde{g}_{\mu\nu}\tilde{g}^{\tau\sigma}\tilde{\nabla}_{\tau}F\tilde{\nabla}_{\sigma}\omega\right)\quad ,
\end{eqnarray}
and the matter energy-momentum tensor $T_{\mu\nu}^{(m)}$ becomes
\begin{equation}
T_{\mu\nu}^{(m)}=\Omega^2\tilde{T}_{\mu\nu}^{(m)}\quad \label{z}.
\end{equation}

There might be an infinite set of representations of physics induced by the conformal transformations due to the ambiguities of $\Omega^2$ \cite{Sante}. However, these representations represent the same physics since they have the same root  in the original Jordan frame which can be seen by doing the inverse transformation. The conformal transformation $\Omega^2$ can be considered as an extra freedom in presenting physics.  If we take $\Omega^2=F$ to fix the freedom associated with the conformal mapping, any modification to the standard gravity in the Jordan frame can have a certain map in the conformal transformation.

Substituting Eqs.~(\ref{G})-(\ref{z}) into Eq.(\ref{p}), we have
\begin{eqnarray}
\tilde{G}_{\mu\nu}&=&\frac{3}{2}\tilde{\nabla}_{\mu}\ln F\tilde{\nabla}_{\nu}\ln F-\frac{3}{4}\tilde{g}_{\mu\nu}\tilde{g}^{\tau\sigma}\tilde{\nabla}_{\tau}\ln F\tilde{\nabla}_{\sigma}\ln F\nonumber\\
&+&\frac{f-RF}{2F^2}\tilde{g}_{\mu\nu}+\kappa^2\tilde{T}_{\mu\nu}^{(m)}\quad \label{q} .
\end{eqnarray}
Defining a scalar field $\varphi$
as $\ln F= \kappa \sqrt{\frac{2}{3}}\varphi$, the energy-momentum tensor for the scalar field reads
\begin{equation}
\tilde{T}_{\mu\nu}^{(e)}=\tilde{\nabla}_{\mu}\varphi\tilde{\nabla}_{\nu}\varphi-\frac{1}{2}\tilde{g}_{\mu\nu}\tilde{g}^{\tau\sigma}\tilde{\nabla}_{\tau}\varphi\tilde{\nabla}_{\sigma}\varphi-\tilde{g}_{\mu\nu}V\quad,
\end{equation}
where the potential $V=\frac{FR-f}{2\kappa^2F^2}$.
From the energy-momentum tensor, we can obtain the Lagrangian density of the field $\varphi$
\begin{equation}
\mathcal{L}_{\varphi}=K-V=-\frac{1}{2}\tilde{g}^{\mu\nu}\tilde{\nabla}_{\mu}\varphi\tilde{\nabla}_{\nu}\varphi-V\quad .
\end{equation}
The kinetic term $K=-\frac{1}{2}\tilde{g}^{\mu\nu}\tilde{\nabla}_{\mu}\varphi\tilde{\nabla}_{\nu}\varphi$ should be positive, which requires
\begin{equation}
\tilde{g}^{\mu\nu}\tilde{\nabla}_{\mu}\varphi\tilde{\nabla}_{\nu}\varphi<0\quad,\label{timelike}
\end{equation}
 meaning that $\tilde{\nabla}_{\nu}\varphi$ is a time-like vector.

The Einstein equation can be rewritten as,
\begin{equation}
\tilde{G}_{\mu\nu}=\kappa^2(\tilde{T}_{\mu\nu}^{(e)}+\tilde{T}_{\mu\nu}^{(m)}).
\end{equation}
This result can also be obtained by conformally transforming the action Eq~(\ref{action}) and then doing the variation \cite{tsujikawa}.

The equation of motion for matter field in the Einstein frame is given by
\begin{equation}
\tilde{\nabla}^{\mu}\tilde{T}_{\mu\nu}^{(m)}=-\frac{\kappa}{\sqrt{6}}\tilde{T}^{(m)}\tilde{\nabla}_{\nu}\varphi
=-\frac{\tilde{T}^{(m)}}{2}\tilde{\nabla}_{\nu}\ln F\quad .
\end{equation}
Recall that $\tilde{\nabla}_{\mu}\varphi$ is time-like and so does $\tilde{\nabla}_{\mu}\ln F$. From the discussion in the appendix, we know that $\tilde{\nabla}_{\mu}\ln F$ relates to the dilation function $\Gamma=\frac{1}{\tilde{m}}\frac{d\tilde{m}}{d\tilde{t}}$ through
\begin{eqnarray}
&&\tilde{g}^{\mu\nu}\tilde{\nabla}_{\mu}\ln F\tilde{\nabla}_{\nu}\ln F=-4\Gamma^2\nonumber\\
&&\tilde{\nabla}_{\mu}\ln F=2\Gamma (\frac{\partial}{\partial \tilde{t}})_{\mu}\quad .\label{dilationgamma}
\end{eqnarray}
where $(\frac{\partial}{\partial \tilde{t}})_{\mu}$ is the normalized four velocity which is parallel to  $\tilde{\nabla}_{\mu}\ln F$.
If $\Gamma=0, F=1$ and $f(R)=R+Constant$. Thus the mass dilation function $\Gamma$ reflects the deviation of the $f(R)$ gravity from the Einstein gravity.

In the following discussion, we will show that the $f(R)$ gravity can be specified by using the mass dilation $\Gamma$. From Eq.~(\ref{otransR}), the Ricci scalar curvature $R$ can be obtained as
\begin{equation}
R=F(\tilde{R}+3\tilde{\Box}\ln F-\frac{3}{2}\tilde{g}^{\mu\nu}\tilde{\nabla}_{\mu}\ln F\tilde{\nabla}_{\nu}\ln F)\quad .\label{transR}
\end{equation}
Employing Eq.(\ref{q}) and substituting (\ref{transR}), we get
\begin{equation}
f=F^2(\frac{\tilde{R}}{2}+3\tilde{g}^{\tau\sigma}\tilde{\nabla}_{\tau}\tilde{\nabla}_{\sigma}\ln F-\frac{3}{4}\tilde{g}^{\tau\mu}\tilde{\nabla}_{\tau}\ln F\tilde{\nabla}_{\mu}\ln F-\frac{\kappa^2}{2}\tilde{T}^{(m)})\quad .\label{f}
\end{equation}

Taking the derivative of Eq.~(\ref{f}), noting
$\tilde{\nabla}_{\mu}f=F\tilde{\nabla}_{\mu}R$ and considering Eq.~(\ref{transR}), we have
\begin{eqnarray}
&&\frac{1}{2}\tilde{\nabla}_{\mu}\tilde{R}+6\Gamma\tilde{\nabla}_{\mu}\Gamma-3\tilde{\nabla}_{\mu}\ln F(\tilde{g}^{\tau\sigma}\tilde{\nabla}_{\tau}\tilde{\nabla}_{\sigma}\ln F)\nonumber\\
&&=-\kappa^2(\tilde{T}^{(m)}\tilde{\nabla}_{\mu}\ln F +\frac{1}{2}\tilde{\nabla}_{\mu}\tilde{T}^{(m)}).\label{R}
\end{eqnarray}

Noting Eq.~(\ref{dilationgamma}),
Eq.~(\ref{R}) now can be rewritten as,
\begin{eqnarray}
\frac{1}{2}\tilde{\nabla}_{\mu}\tilde{R}+6\Gamma\tilde{\nabla}_{\mu}\Gamma-12\Gamma (\frac{d\Gamma}{d\tilde{t}}+\Gamma\tilde{\theta})(\frac{\partial}{\partial \tilde{t}})_{\mu}\nonumber \\
=-\kappa^2(2\tilde{T}^{(m)} \Gamma(\frac{\partial}{\partial \tilde{t}})_{\mu} +\frac{1}{2}\tilde{\nabla}_{\mu}\tilde{T}^{(m)})\label{scalarR}.
\end{eqnarray}
For $\Gamma=0$, this is just the equation in Einstein gravity.
In Eq.(\ref{scalarR}), $\frac{d\Gamma}{d\tilde{t}}=(\frac{\partial}{\partial \tilde{t}})_{\sigma}\tilde{g}^{\sigma \tau}\tilde{\nabla}_{\tau}\Gamma$ and  $\tilde{\theta}=\tilde{g}^{\sigma \tau}\tilde{\nabla}_{\tau}(\frac{\partial}{\partial \tilde{t}})_{\sigma}$ is the expansion function
satisfying the Raychaudhuri  equation
\begin{eqnarray}
\frac{d\tilde{\theta}}{d\tilde{t}}&=&-\frac{1}{3}\tilde{\theta}^2-\tilde{\sigma}_{\mu\nu}\tilde{\sigma}^{\nu\mu}+\tilde{\omega}_{\mu\nu}\tilde{\omega}^{\mu\nu}+{\tilde{\rm D}}^{\tau}\tilde{A}_{\tau}\nonumber\\
&+&\tilde{A}^{\tau}\tilde{A}_{\tau}-\tilde{R}_{\mu\nu}(\frac{\partial}{\partial \tilde{t}})^{\mu}(\frac{\partial}{\partial \tilde{t}})^{\nu},\label{Ray}
\end{eqnarray}
where $\tilde{A}^{\tau}$ is the four acceleration, $\omega_{\mu\nu}={\rm \tilde{D}}_{[\nu}(\frac{\partial}{\partial \tilde{t}})_{\mu]}$ is the vorticity tensor
, $\sigma_{\mu\nu}={\rm \tilde{D}}_{<\nu}(\frac{\partial}{\partial \tilde{t}})_{\mu>}$ is the shear tensor, ${\rm \tilde{D}}$ is spatial derivatives defined as
\begin{equation}
{\tilde{\rm D}}_{e}{\tilde{S}_{ab\cdots}}^{cd\cdots}=\tilde{h}_{e}^{s}\tilde{h}_{a}^{f}\tilde{h}_{b}^{g}\tilde{h}_{q}^{c}\cdots\tilde{\nabla}_s{\tilde{S}_{fg\cdots}}^{qr\cdots}
\end{equation}
for arbitrary tensor ${\tilde{S}_{ab\cdots}}^{cd\cdots}$ field. $\tilde{h}_{\mu\nu}=\tilde{g}_{\mu\nu}+(\frac{\partial}{\partial \tilde{t}})_{\mu}(\frac{\partial}{\partial \tilde{t}})_{\nu}$ is the projection operator.

Taking $3+1$ decomposition of Eq.~(\ref{scalarR}), the time-like part reads,
\begin{equation}
\frac{1}{2}\frac{d\tilde{R}}{d\tilde{t}}+18\Gamma\frac{d\Gamma}{d\tilde{t}}+12\Gamma^2\tilde{\theta}=\kappa^2(2\tilde{T}^{(m)} \Gamma -\frac{1}{2}\frac{d\tilde{T}^{(m)}}{d\tilde{t}})\label{Rtime}\quad,
\end{equation}
and
the spatial part is
\begin{equation}
\frac{1}{2}\tilde{\rm{D}}_{\mu}\tilde{R}+6\Gamma\tilde{\rm{D}}_{\mu}\Gamma=-\frac{\kappa^2}{2}\tilde{\rm{D}}_{\mu}\tilde{T}^{(m)}\label{Rspace}\quad.
\end{equation}

From the above two equations, it is clear that the dilation function $\Gamma$ marks the deviation of the $f(R)$ gravity from the Einstein gravity. Once  the
dilation function is specified, $f(R)$ gravity can be completely constructed. In the next section, we will apply this formalism to describe the late time acceleration in cosmology.

\section{late time acceleration in $f(R)$ gravity \label{II}}
\subsection{The Einstein frame}
In the Einstein frame, the flat Friedmann-Robertson-Walker (FRW) line element reads
\begin{equation}
d\tilde{s}^2 = -d\tilde{t}^2+\tilde{a}^2d\tilde{x}^2\quad .
\end{equation}
The expansion observed by the comoving observer is
$\tilde{\theta}=3\tilde{H}$,
where the Hubble parameter $\tilde{H}$ is defined by
$\tilde{H}=\frac{d\tilde{a}}{d\tilde{t}}/\tilde{a}$. For the perfect fluid and the comoving observer, we have $\tilde{\sigma}_{\mu\nu}=\tilde{\omega}_{\mu\nu}=\tilde{A}_{\tau}=0$, so that Eq. ~(\ref{Ray}) reduces to
\begin{equation}
\frac{d\tilde{H}}{d\tilde{x}}=\frac{\tilde{R}}{6\tilde{H}}-2\tilde{H}\label{1}\quad ,
\end{equation}
where $\tilde{x}=\ln \tilde{a}$, $\tilde{R}=-2\tilde{R}_{00}+6\tilde{H}^2$ and $\tilde{R}=6(2\tilde{H}^2+\frac{d\tilde{H}}{d\tilde{t}})$.

From Eq.~(\ref{Rtime}), we find
\begin{equation}
\frac{d\tilde{R}}{d\tilde{x}}+36\Gamma\frac{d\Gamma}{d\tilde{x}}+72\Gamma^2=-3\kappa^2\tilde{\rho}_m(\frac{\Gamma}{\tilde{H}}+1)\label{2}\quad ,
\end{equation}
where we have neglected the pressure of the matter so that $\tilde{T}^{(m)}=-\tilde{\rho}_m$. After the conformal transformation, the mass is no longer conserved. It satisfies the continuity equation
\begin{equation}
\frac{d\tilde{\rho}_m}{d\tilde{x}}+3\tilde{\rho}_m=\frac{\Gamma}{\tilde{H}}\tilde{\rho}_m\label{3}.
\end{equation}
From the relation between $F$ and $\Gamma$ and expressing $F$ into $\varphi$, we have the continuity equation for the scalar field
\begin{equation}
\frac{d\varphi}{d\tilde{x}}=-\frac{\sqrt{6}\Gamma}{\kappa \tilde{H}}.\label{4}
\end{equation}
For a homogeneous universe, Eq.~(\ref{Rspace}) is automatically satisfied.
Once the dilation function is specified, we can explore the expansion history of the universe in the Einstein frame by solving Eqs.~(\ref{1})-(\ref{4}) with proper boundary conditions.

 Before proceeding, we discuss the intrinsic consistency of the formalism described above  with the usual interacting model with a scalar field.
From Eq.~(\ref{f}), $f$ can be obtained in terms of $\Gamma$
\begin{equation}
f=F^2(\frac{\tilde{R}}{2}+3\Gamma^2+6\tilde{H}\frac{d\Gamma}{d\tilde{x}}+18\Gamma\tilde{H}+\frac{1}{2}\kappa^2\tilde{\rho}_m)\quad ,\label{trf}
\end{equation}
where $\ln F =\kappa \sqrt{\frac{2}{3}}\varphi$ has been employed. Furthermore, from Eq.~(\ref{transR}), we have
\begin{equation}
R=F\tilde{R}+F(6\Gamma^2+6\tilde{H}\frac{d\Gamma}{d\tilde{x}}+18\Gamma\tilde{H}).~\label{trR}
\end{equation}
Noting $\frac{d\ln F}{d\tilde{t}}=-2\Gamma$, $\frac{d}{d\tilde{t}}=\tilde{H}\frac{d}{d\tilde{x}}$ and combining Eqs~(\ref{trf},\ref{trR}), we get
\begin{equation}
\tilde{H}^2=\frac{\kappa^2}{3}\left[\tilde{\rho}_m+\frac{1}{2}\left(\frac{d\varphi}{d\tilde{t}}\right)^2+\frac{FR-f}{2\kappa^2F^2}\right]\quad \label{fredmannEinstein} ,
\end{equation}
which is exactly the Friedmann equation if we define energy density and the pressure of the scalar field as
\begin{eqnarray}
\tilde{\rho}_d&=&\frac{1}{2}\left(\frac{d\varphi}{d\tilde{t}}\right)^2+V=\frac{3\Gamma^2}{\kappa^2}+V\nonumber \\
\tilde{p}_d&=&\frac{1}{2}\left(\frac{d\varphi}{d\tilde{t}}\right)^2-V=\frac{3\Gamma^2}{\kappa^2}-V\quad, \label{density}
\end{eqnarray}
where
\begin{eqnarray}
V&=&\frac{FR-f}{2\kappa^2F^2}\quad .
\end{eqnarray}

If the scalar field plays the role of DE in the Einstein frame, its continuity equation reads
\begin{equation}
\frac{d\tilde{\rho}_d}{d\tilde{x}}+3(1+\tilde{w})\tilde{\rho}_d=-\frac{\Gamma}{\tilde{H}}\tilde{\rho}_m\quad ,
\end{equation}
where $\tilde{w}=\tilde{p}_d/\tilde{\rho}_d$ is the equation of state of DE. This leads to the equation of motion of the scalar field
\begin{equation}
\frac{d^2\varphi}{d\tilde{t}^2}+3\tilde{H}\frac{d\varphi}{d\tilde{t}}+\frac{\partial V}{\partial \varphi}=\frac{\kappa}{\sqrt{6}}\tilde{\rho}_m \quad .
\end{equation}

The analysis above shows that, in the Einstein frame,
the $f(R)$ cosmology is conformally equivalent to the model of interaction between
dark sectors. There is a freedom in choosing the coupling strength $\Gamma$. However, it must be consistent with the viability condition of the $f(R)$ gravity.

A viable $f(R)$ model must pass the stringent local test, which requires that in the dense region the model should go back to the standard Einstein gravity,
$\lim_{\tilde{R}\longrightarrow\infty}\Gamma=0$.
On the other hand, in the lower dense region $f(R)$ gravity should have enough deviation from the Einstein gravity to achieve the late time acceleration of the universe. However, in the vacuum, the dilation should not go to infinity,
$\lim_{\tilde{R}\longrightarrow0}\Gamma<\infty$.
In Eq.~(\ref{dilation}), $\Gamma$ is defined as the dilation rate of the mass of test particles in gravitational fields. To satisfy the weak equivalent principle, $\Gamma$ should be independent of the species of matter \cite{Fujii}. One natural choice of the $\Gamma$ form is to consider it as a geometris quantity, a function of $\tilde{R}$. There is quite a wide range of the choice for $\Gamma$ which can satisfy $\lim_{\tilde{R}\longrightarrow\infty}\Gamma=0$ and $\lim_{\tilde{R}\longrightarrow0}\Gamma<\infty$ to pass the local test and have reasonable expansion history. In this work, we only present one of the most simplest choices. We take $\Gamma$ as
\begin{equation}
\Gamma=\frac{\alpha}{ \tilde{R}+\beta}\quad ,
\end{equation}
which satisfies $\lim_{\tilde{R}\longrightarrow\infty}\Gamma=0$ and $
\lim_{\tilde{R}\longrightarrow0}\Gamma=\frac{\alpha}{\beta}$
where $\alpha$ and $\beta$ are constants.

\begin{figure*}
\centering
\includegraphics[width=6in,height=6in]{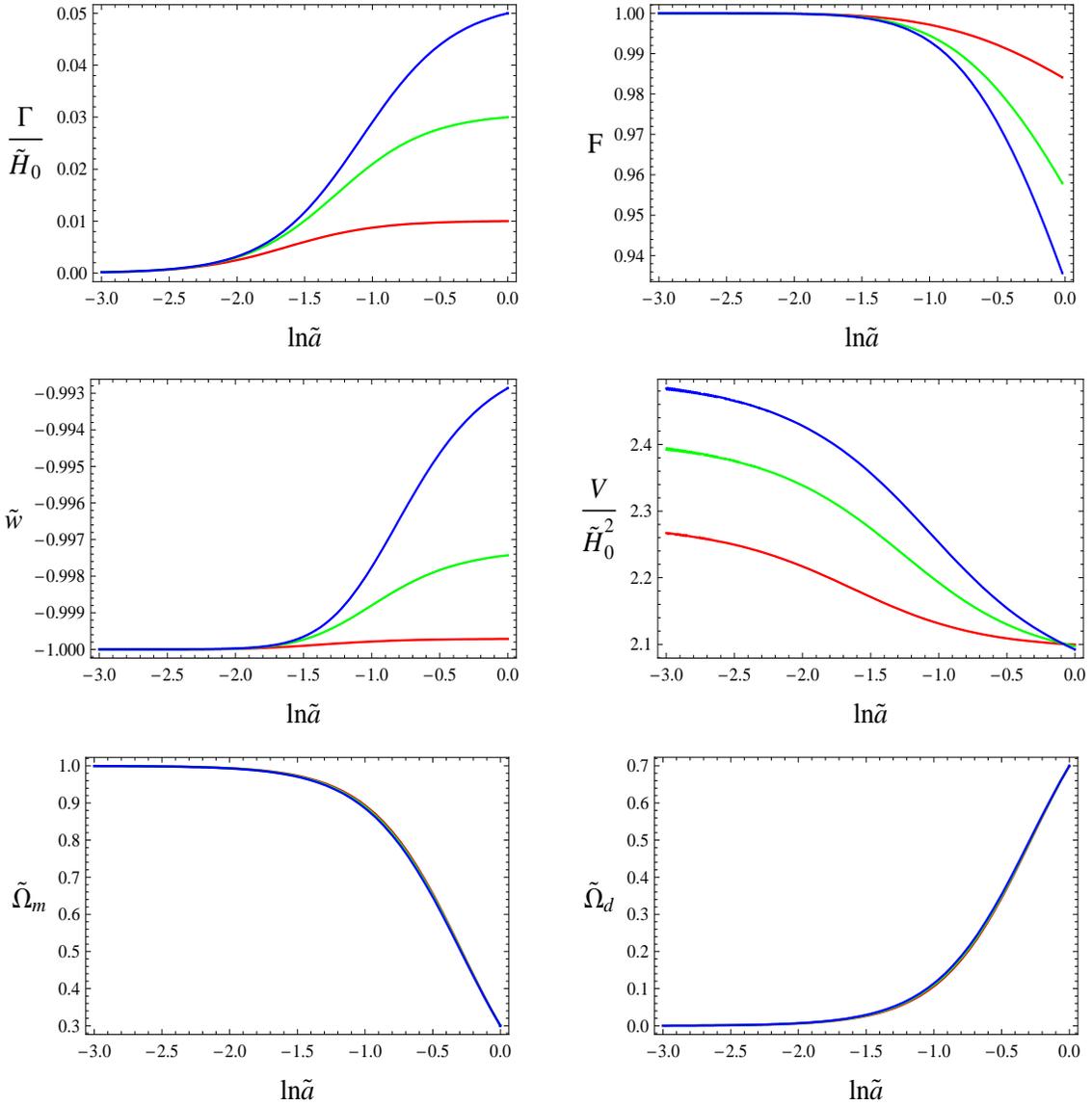}
\caption{Viable description of the universe by $f(R)$ gravity in the Einstein frame. We set the cosmological parameters as $\tilde{\Omega}_m^0=0.3,\tilde{\Omega}_d^0=0.7,\alpha/\tilde{H}_0^3=1.2$ at the present moment. The red, green and blue curves represent the models for $\Gamma_0/\tilde{H}_0=0.01$,$\Gamma_0/\tilde{H}_0=0.03$, and $\Gamma_0/\tilde{H}_0=0.05$ respectively. The values of the $\beta$-parameter
 are, respectively, $\beta/\tilde{H}_0^2=110.702,\beta/\tilde{H}_0^2=30.7162,\beta/\tilde{H}_0^2=14.745$. }\label{Einsteinframe}
\end{figure*}

With the specified $\Gamma$, we can obtain the evolution of the universe by solving Eqs.~(\ref{1})-(\ref{4}) in the Einstein frame. This is equivalent to solving the model of interaction between DE and DM. For convenience $\tilde{\rho}_m $, $\tilde{R}$ are solved in the unit $\tilde{H}_0^2$ and $\Gamma$ in the unit $\tilde{H}_0$. $\alpha$ and $\beta$ are set in units $\tilde{H}_0^3$ and $\tilde{H}_0^2$ respectively. We set the starting point at present,
$\tilde{x}=0$, with the initial conditions $\tilde{R}_0=12-3\tilde{\rho}_m^0-18\Gamma_0^2$, $\tilde{\rho}_m^0=3\tilde{\Omega}_m^0$ and $\tilde{H}_0=1$. In order to let our model fully return the standard Einstein gravity in the past, we need $F\rightarrow 1$, which is equivalent to setting the boundary condition
$\lim_{\tilde{x}\longrightarrow-\infty}\varphi\rightarrow0$. The numerical results are shown in Fig~\ref{Einsteinframe}.

We see that in the early universe, when the $f(R)$ gravity boils down to Einstein gravity, the dilation $\Gamma$ disappears. It becomes nonzero only when the theory deviates from the Einstein gravity. The larger the deviation, the bigger shall be $\Gamma$. In the language of the interacting model, we see that the strength of the interaction becomes stronger in the late time universe, while in the early time, DE and DM evolves independently, which provides a mechanism to recover the standard radiation and matter dominated phase in the universe expansion history.

\subsection{The Jordan frame}

In order to further examine the consistency between the $f(R)$ gravity and the model of interaction between dark sectors, let's go back to the Jordan frame.
Noting that
\begin{equation}
d\tilde{t}=\Omega dt,\quad,d\tilde{r}=\Omega dr,\quad, \tilde{a}=\Omega a,
\end{equation}
the Hubble expansion in the Jordan frame reads,
\begin{equation}
H=\Omega \tilde{H}-\frac{d\Omega}{d\tilde{t}}=F^{1/2}(\tilde{H}-\frac{1}{2}\frac{d \ln F}{d\tilde{t}}),\label{Hubblejordan}
\end{equation}
and
\begin{equation}
H^2=F\left[\tilde{H}^2+\frac{1}{4}\left(\frac{d\ln F}{d\tilde{t}}\right)^2-\tilde{H}\frac{d\ln F}{d\tilde{t}}\right].\label{Hsquare}
\end{equation}
Employing Eq.(\ref{fredmannEinstein}), the Friedmann equation in the Jordan frame becomes
\begin{equation}
H^2=\frac{FR-f}{6F}-H\frac{\dot{F}}{F}+\frac{\kappa^2}{3F}\rho_m\quad,\label{Fredmann}
\end{equation}
where $\rho_m=F^2\tilde{\rho}_m$ and
the Ricci scalar field $R$ reads
\begin{eqnarray}
R&=&6(2H^2+\dot{H})\nonumber \\
&=&F6(2\tilde{H}^2+\frac{d\tilde{H}}{d\tilde{t}})\nonumber \\
&+&F\left[\frac{3}{2}\left(\frac{d \ln F}{d\tilde{t}}\right)^2-9\frac{d \ln F}{d\tilde{t}}\tilde{H}-3\frac{d^2 \ln F}{d\tilde{t}^2}\right]\label{trRJordan}\quad,
\end{eqnarray}
with
\begin{equation}
\frac{dH}{dt}=\frac{F}{2}\frac{d\ln F}{d\tilde{t}}\tilde{H}+F\frac{d\tilde{H}}{d\tilde{t}}-\frac{F}{2}\frac{d^2\ln F}{d\tilde{t}^2}-\frac{F}{4}\left(\frac{d \ln F}{d\tilde{t}}\right)^2.\nonumber
\end{equation}
The dot denotes the derivative with respect to $t$. The above Friedmann equation can also be derived directly from Eq. (\ref{Einstein}).

The Friedmann equation Eq.(\ref{Fredmann}) can be recast into the form
\begin{eqnarray}
&&y''-(1+\frac{E'}{2E}+\frac{4E''+E'''}{4E'+E''})y'+\frac{4E'+E''}{2E}y\nonumber\\
&=&\Omega_m^0e^{-3x}\frac{3(4E'+E'')}{E},\label{Eqy}
\end{eqnarray}
where $E=\frac{H^2}{H_0^2}, y=\frac{f}{H_0^2}$. In the derivation, we have used the relation
$F=\frac{f'}{R'}$,
where prime denotes $\frac{d}{dx}$ and $x=\ln a$, and
$F'=\frac{f''}{R'}-\frac{R''f'}{(R')^2}$ with
$R=3\left[4H^2+(H^2)'\right]$.
Substituting $f$ by $f-R$, Eq~(\ref{Eqy}) goes back to the
result obtained in ~\cite{Song}.

The background expansion $E$ can be parameterized as
\begin{equation}
 E=(1-\Omega^0_d)e^{-3x}+\Omega^0_de^{-3(1+w)x}\quad\label{HubbleE} .
\end{equation}
However this $E$ cannot fully fix the $f(R)$ model.
The $f(R)$ model has an external choice of the $f$ form which
 satisfies the differential equation of Eq~(\ref{Eqy}). However, the boundary conditions for $y,y'$ are not completely free because a viable $f(R)$ model should pass the local test
\begin{equation}
\lim_{R\rightarrow\infty}f(R)/R=\lim_{R\rightarrow\infty}\frac{\frac{\partial f(R)}{\partial R}}{\frac{\partial R}{\partial R}}=\lim_{R\rightarrow\infty}F\rightarrow1\label{boundary}\quad,
\end{equation}
which puts  constraints on $y_0$ and $y'_0$. There is a freedom in $f(R)$ models, which
can be represented by a dimensionless quantity\cite{Song}
\begin{equation}
B=\frac{f_{RR}}{F}R'\frac{H}{H'}=\frac{d \ln F}{d\ln H}\quad.
\end{equation}
The quantity $B$ relates to $\Gamma$ in the form
\begin{equation}
B=-2\Gamma\frac{\sqrt{F}}{H'}\quad.\label{zz}
\end{equation}
From Eq~(\ref{boundary}), $B$ satisfies
$\lim_{R\rightarrow\infty}B\rightarrow 0$, which imposes the constraint
 $$\lim_{R\rightarrow\infty}\Gamma\rightarrow 0\quad.$$ The evolution of $B$ is not entirely free. Only the boundary value $B_0$ is a free parameter, which characterizes the $f(R)$ model \cite{Song}.

\begin{figure*}
\centering
\includegraphics[width=6in,height=6in]{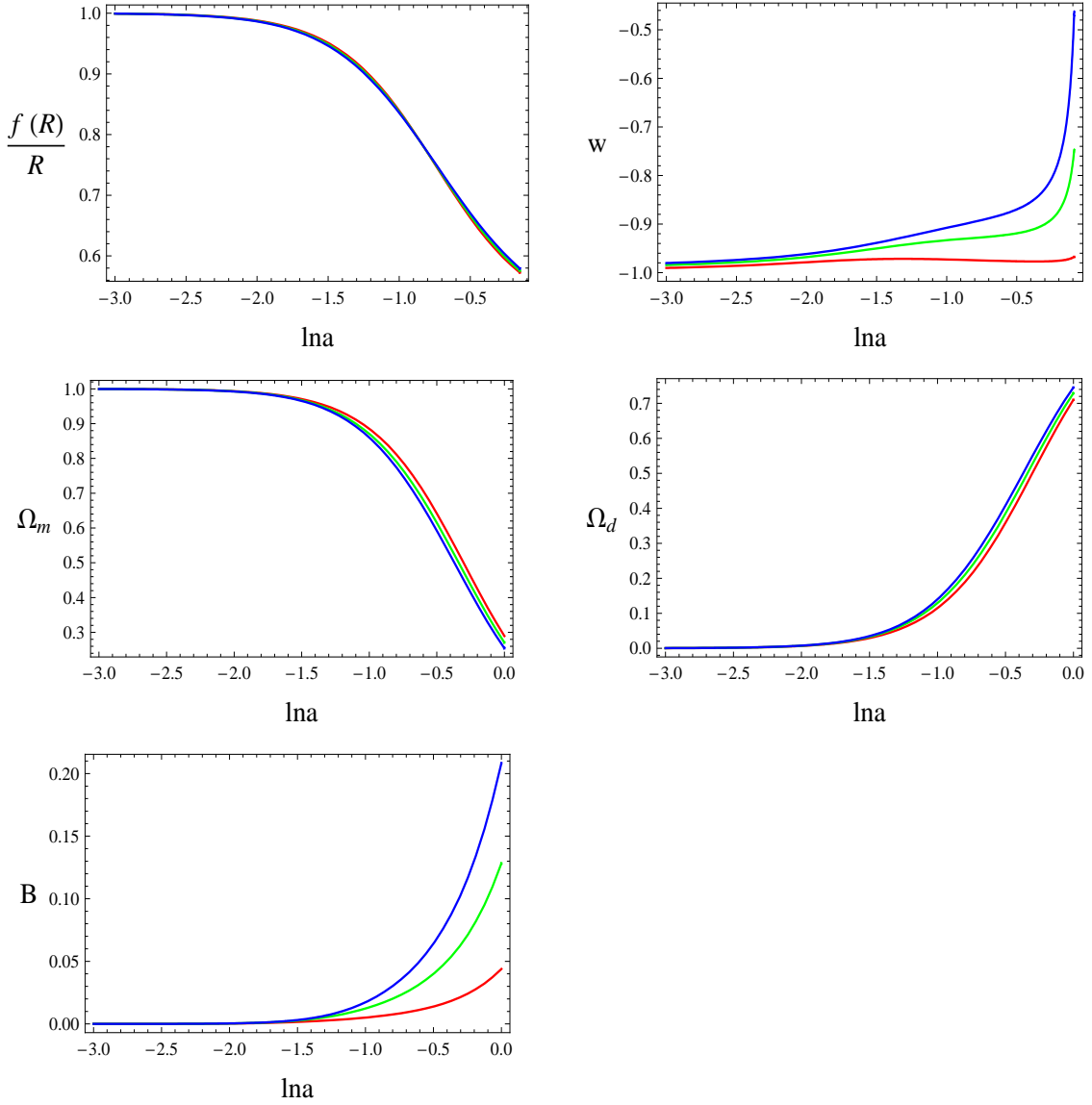}
\caption{The $f(R)$ cosmology in Jordan frame obtain from the models in Einstein frame by using the inverse conformal transformation}\label{Jordanframe}
\end{figure*}

In the cosmological scale, it was argued that the $f(R)$ gravity can reduce to the large-scale CMB anisotropy and avoid the high curvature instability only when $B>0$ \cite{Song}. Considering the Hubble expansion Eq.~(\ref{Hubblejordan}), we can use the inverse conformal transformation to get $H$ and in combination with Eq.~(\ref{HubbleE}), we can figure out the effective DE equation of state $w$ . Similarly, using the inverse map of the conformal transformation, we can obtain other quantities in the Jordan frame from the Einstein frame. The numerical results are shown in Fig.\ref{Jordanframe}. From the results, we find that $w>-1$ and this ensures $H'<0$. Thus from Eq~(\ref{zz}) we learn that the condition $B>0$ leads to $\Gamma>0$. This is an interesting observation, since it shows that the viability condition of the $f(R)$ gravity in the cosmological scale in Jordan frame urges the coupling between dark sectors to be positive in Einstein frame. From Eq~(\ref{3}), the positive coupling indicates that the energy flows from DE to DM, which is the requirement to diminish the coincidence problem \cite{19}\cite{199}. Thus the viable $f(R)$ gravity in the cosmological scale has a predisposition to alleviate the coincidence problem in the conformal theory of interaction between dark sectors.

\section{CONCLUSIONS\label{III}}

In this paper, we have examined the conformal equivalence between the $f(R)$ gravity and the interacting dark sector model in the Einstein frame. We construct the $f(R)$ model in terms of the mass dilation. Once the dilation function is known, the $f(R)$ gravity can be constructed. Studying cosmology in the Einstein frame, the $f(R)$ gravity is conformally equivalent to the interacting model. The strength of the interaction marks the deviation of the $f(R)$ gravity from the Einstein gravity. In the earlier time of the universe, there is no interaction in the dark sector in the Einstein frame and the $f(R)$ gravity returns to the Einstein gravity. In the later epoch the interaction becomes stronger and the deviation of the $f(R)$ gravity from the Einstein theory is larger. More interestingly we found that the condition that the $f(R)$ gravity agrees with the CMB observation requires the energy to flow from DE to DM in the corresponding interacting model. This is actually consistent with the requirement to alleviate the coincidence problem in the Einstein framework. This further shows the equivalence between the $f(R)$ gravity and the interaction model.

\section{Appendix }
\subsection{ Conformal transformation}
We briefly summarize the basic properties of conformal transformation. The metric changes
\begin{equation}
\tilde{ds}^2=\Omega^2ds^2,\quad,\tilde{g}_{ab}=\Omega^2g_{ab},\quad \tilde{g}^{ab}=g^{ab}/\Omega^2.
\end{equation}
The difference of the covariant derivative between $\nabla_{\mu}$ and $\tilde{\nabla}_{\mu}$ is characterized by  ${C^{\tau}}_{\mu\nu}$,
\begin{equation}
\nabla_{\mu}A_{\nu}=\tilde{\nabla}_{\mu}A_{\nu}+{C^{\tau}}_{\mu\nu}A_{\tau}\quad,
\end{equation}
where $A_{\nu}$ is an arbitrary vector and ${C^{\tau}}_{\mu\nu}$ is given by
\begin{equation}
{C^{\tau}}_{\mu\nu}=2{\delta^{\tau}}_{(\mu}\tilde{\nabla}_{\nu)}\omega-\tilde{g}_{\mu\nu}\tilde{g}^{\tau\sigma}\tilde{\nabla}_{\sigma}\omega\quad ,
\end{equation}
where $\omega=\ln \Omega$.

For the scalar field $\varphi$, the transformation with second order derivative satisfies
\begin{equation}
\nabla_{\mu}\nabla_{\nu}\varphi=\tilde{\nabla}_{\mu}\tilde{\nabla}_{\nu}\varphi+2\tilde{\nabla}_{(\mu}\varphi\tilde{\nabla}_{\nu)}\omega- \tilde{g}_{\mu\nu}\tilde{g}^{\tau\sigma}\tilde{\nabla}_{\tau}\varphi\tilde{\nabla}_{\sigma}\omega\quad .
\end{equation}

The conformal transformation for the Ricci tensor obeys
\begin{eqnarray}
R_{\mu\nu}&=&\tilde{R}_{\mu\nu}+2\tilde{\nabla}_{\mu}\omega\tilde{\nabla}_{\nu}\omega+2\tilde{\nabla}_{\mu}\tilde{\nabla}_{\nu}\omega+\tilde{g}_{\mu\nu}\tilde{\Box } \omega\nonumber\\
&-&2\tilde{g}_{\mu\nu}\tilde{g}^{\tau\sigma}\tilde{\nabla}_{\tau}\omega\tilde{\nabla}_{\sigma}\omega\quad ,
\end{eqnarray}
where $
\tilde{\Box }=\tilde{g}^{\mu\nu}\tilde{\nabla}_{\mu}\tilde{\nabla}_{\nu}
$.
The scalar Ricci curvature reads
\begin{equation}
R=\Omega^2(\tilde{R}+6\tilde{\Box}\omega-6\tilde{g}^{\mu\nu}\tilde{\nabla}_{\mu}\omega\tilde{\nabla}_{\nu}\omega)\quad \label{otransR}.
\end{equation}

The differential operator $d$ does not change under such transformation,
$
d=\tilde{d}
$.
If $d$ acts on a scalar field $f$, we find that
\begin{equation}
\partial_{\mu}f=(df)_{\mu}=(\tilde{d}f)_{\mu}=\tilde{\partial}_{\mu}f.
\end{equation}
\subsection{Physical interpretation of the conformal transformation}
The conformal transformation rescales the basic units used in the original frame \cite{dicke}.
\begin{equation}
d\tilde{t}=\Omega dt,\quad,d\tilde{r}=\Omega dr
\end{equation}
where $dr$,$dt$ denote the space and time separation. Although the conformal factor $\Omega$ is arbitrary,  the basic physics does not change under the conformal transformation.

The equation of motion under such a transformation becomes
\begin{equation}
\tilde{\nabla}_{\mu}\tilde{T}^{\mu \nu}=\tilde{\nabla}_{\mu}(\frac{1}{\Omega^6}T^{\mu \nu})=-\tilde{T}\tilde{g}^{\mu\nu}\partial_{\mu}\ln\Omega=-\tilde{T}\frac{\tilde{\partial}^{\nu}\Omega}{\Omega}\quad \label{efm},
\end{equation}
where $\tilde{T}^{\mu \nu}\equiv\tilde{g}^{\mu\tau}\tilde{g}^{\nu\sigma}\tilde{T}_{\tau \sigma}$, $T^{\mu \nu}\equiv g^{\mu\tau}g^{\nu\sigma}T_{\tau \sigma}$ and
\begin{equation}
\tilde{T}_{\mu\nu}=\frac{1}{\Omega^2}T_{\mu\nu}\quad,\tilde{T}=\tilde{g}^{\mu\nu}\tilde{T}_{\mu\nu}=\frac{g^{\mu\nu}}{\Omega^4}T_{\mu\nu}=\frac{T}{\Omega^4}.\label{tensortrans}
\end{equation}
We have used Eq.(D.20) in the appendix of \cite{wald} in deriving (\ref{efm}).

For a single particle,  Eq~(\ref{efm}) reduces to
\begin{equation}
\tilde{u}^{\mu}\tilde{\nabla}_{\mu}\tilde{p}^{\nu}=\frac{\tilde{m}}{\Omega}\tilde{\partial}^{\nu}\Omega\quad \label{EQM}.
\end{equation}
$\tilde{m}\tilde{\partial}^{\nu}\ln \Omega$ is the interaction vector which can be either timelike or spacelike. However, in this work, we only focus on the case when $\tilde{\partial}^{\nu}\ln \Omega$ is timelike, because the interaction induced by modified gravity is always in this case Eq~(\ref{timelike}).

Eq~(\ref{EQM}) is consistent with the well-established physics, if comparing with the equation of motion of particles with varying mass in general relativity \cite{He,199}
\begin{equation}
\tilde{u}^{\mu}{\tilde{\nabla}}_{\mu}\tilde{p}^{\nu}=\frac{d\tilde{m}}{d\tilde{t}}(\frac{\partial}{\partial\tilde{t}})^{\nu}\quad ,\label{test}
\end{equation}
where $(\frac{\partial}{\partial\tilde{t}})^{\nu}(\frac{\partial}{\partial\tilde{t}})_{\nu}=\tilde{g}_{\mu\nu}(\frac{\partial}{\partial\tilde{t}})^{\mu}(\frac{\partial}{\partial\tilde{t}})^{\nu}=-1$.

We obtain
\begin{equation}
\frac{\tilde{m}}{\Omega}\tilde{\partial}^{\nu}\Omega=\frac{d\tilde{m}}{d\tilde{t}}(\frac{\partial}{\partial\tilde{t}})^{\nu}\quad .
\end{equation}
Contracting  the above equation with $(\frac{\partial}{\partial \tilde{t}})_{\nu}$, we arrive at
\begin{equation}
\frac{d\ln \tilde{m}}{d\tilde{t}}=\frac{d \ln\Omega^{-1}}{d\tilde{t}},
\end{equation}
where $\tilde{m}=m/\Omega$ and $m$ is a constant. When $\Omega=1$, $m$ is the mass measured in the original frame.

Suppose a system has the volume $V$ and is composed of $N$ particles, the number  density of particles is defined by $n=N/V$. Considering $d\tilde{t}=\Omega dt, d\tilde{r}=\Omega dr, \tilde{V}=\Omega^3V , \tilde{m}=m/\Omega$ \cite{dicke}, we have
$\tilde{n}=\frac{1}{\Omega^3}n\quad .$
Thus after the conformal transformation, the energy density becomes
$
\tilde{\rho}=\tilde{m}\tilde{n}=\frac{1}{\Omega^4}mn=\frac{1}{\Omega^4}\rho,
$
which is consistent with Eq~(\ref{tensortrans}).

We can introduce a scalar field $\Gamma$ which satisfies
\begin{equation}
\tilde{\rho}_m\frac{\tilde{\partial}^{\mu}\Omega}{\Omega}=\Gamma\tilde{\rho}_m(\frac{\partial}{\partial \tilde{t}})^{\mu}\quad .
\end{equation}
This $\Gamma$ can be rewritten as
\begin{equation}
\Gamma=-\frac{d\ln \Omega}{d\tilde{t}}=\frac{1}{\tilde{m}}\frac{d\tilde{m}}{d\tilde{t}}=\frac{1}{m}\frac{d\tilde{m}}{dt}\quad \label{dilation},
\end{equation}
which clearly indicates the mass dilation rate due to the conformal transformation.

In order to see more clearly on the physical meaning of $\Gamma$, we take the 3+1 decomposition of Eq.~(\ref{test}) relative to a timelike observer $\tilde{Z}^{\mu}=(\frac{\partial}{\partial\tilde{s}})^{\mu}$
\begin{eqnarray}
&&\frac{d\tilde{E}}{d\tilde{\tau}}=\frac{d}{d\tilde{\tau}}(\tilde{\gamma}\tilde{m})=\tilde{\gamma}_{t}\frac{d\tilde{m}}{d\tilde{t}}=\tilde{\gamma}_{t}\Gamma\tilde{m}\nonumber\\
&&\frac{d\tilde{p}^i}{d\tilde{\tau}}=\frac{d}{d\tilde{\tau}}(\tilde{\gamma}\tilde{m}\tilde{v}^i)=\tilde{\gamma}_{t}\frac{d\tilde{m}}{d\tilde{t}}\tilde{v}^i_t=\tilde{\gamma}_{t}\Gamma\tilde{m}\tilde{v}^i_t
\end{eqnarray}
where $\tilde{E}$ and $\tilde{p}^i$ are local energy and three momentum measured by observer $\tilde{Z}^{\mu}$ respectively, $\tilde{u}^{\mu}=(\frac{d}{d\tilde{\tau}})^{\mu}$ is the four velocity of test particle, $\tilde{\gamma}\equiv\frac{d\tilde{s}}{d\tilde{\tau}}$, $\tilde{\gamma}_t\equiv\frac{d\tilde{s}}{d\tilde{t}}$ are Lorentz boot factors, $\tilde{v}^i=d\tilde{x^i}/d\tilde{s}$ is three velocity observed by $\tilde{Z}^{\mu}$. The mass dilation will introduce some external terms on the right hand side of the equation of motion. The spatial part $\tilde{\gamma}_{t}\Gamma\tilde{m}\tilde{v}^i_t$ is an external three force  induced by the energy (mass) change in the test particle. This effect is well known as ``Doppler effect" induced by the ``inertia of energy". However, in cosmology, $\tilde{\gamma}_{t}\Gamma\tilde{m}\tilde{v}^i_t$ is usually called the fifth force.
The fifth force $\tilde{\gamma}_{t}\Gamma\tilde{m}\tilde{v}^i_t$ depends on the choice of the observer. For instance, if we take the observer $\tilde{Z}^{\mu}=(\frac{\partial}{\partial\tilde{s}})^{\mu}$ parallel to $\tilde{\partial}^{\nu}\ln\Omega$, $\tilde{\gamma}_{t}\Gamma\tilde{m}\tilde{v}^i_t$ completely vanishes. However, the local energy $\tilde{E}$ is not conserved if the interaction is induced by gravity. This is because the energy is not localized in the gravitational field. The local measurement of energy $\tilde{E}$ depends on the spacetime. Thus the effect of the timelike part always exists and the interaction vector $\tilde{\partial}^{\nu}\ln \Omega$ is necessarily timelike. The mass dilation is a dominated effect due to the fact that $\tilde{v}^i_t<<1$. Comparing with the timelike part effect, the spacelike fifth force can be neglected.

Moreover, if the dilation rate $\Gamma$ is independent of the species of matter, the equation of motion is only determined by geometry, which means that the weak equivalence principle still holds \cite{Fujii}.

\emph{Acknowledgment: }This work was partially supported by NNSFC and the Ministry of Science and Technology of China.  E. A. acknowledges the support by FAPESP and CNPQ, Brazil.

\end{document}